\documentclass[12pt]{article}
\usepackage[dvips]{graphicx}
\usepackage{amsthm}
\usepackage{epsfig}  
\usepackage{amssymb}

\def\be{\begin{equation}}
\def\bea{\begin{eqnarray}}
\def\ee{\end{equation}}
\def\eea{\end{eqnarray}}

\def\pt{\partial}

\def\dt{\delta}
\def\Dt{\Delta}
\def\eps{\varepsilon}
\def\ffi{\varphi}

\def\La{\Lambda}

\def\om{\omega}

\def\dd{\mbox{d}}

\def\calH{{\cal H}}
\def\calP{{\cal P}}

\begin{document}
\begin{center}
{\Large \bf Surfatron acceleration of a relativistic particle by electromagnetic plane wave}\\

\vskip 10mm

{\Large  A. I. Neishtadt$^{1,2}$, A.~A.~Vasiliev$^{1}$, \\ and A. V. Artemyev$^{1}$ }\\

\vskip 5mm

{\large \it $^{1}$ Space Research Institute, Moscow, Russia}\\
{\large \it $^{2}$ Department of Mathematical Sciences, \\ Loughborough University, UK}\\

\end{center}

\section*{Abstract}

We study motion of a relativistic charged particle in a plane slow electromagnetic wave and background uniform magnetic field. The wave propagates normally to the background field. Under certain conditions, the resonance between the wave and the Larmor motion of the particle is possible. Capture into this resonance results in acceleration of the particle along the wave front (surfatron acceleration). We analyse the phenomenon of capture and show that a captured particle never leaves the resonance and its energy infinitely grows. Scattering on the resonance is also studied. We find that this scattering results in diffusive growth of the particle energy. Finally, we estimate energy losses due to radiation by an accelerated particle.

\section{ Introduction}\label{intro}

Motion of a charged particle in an inhomogeneous     electromagnetic field can be described by a Hamiltonian nonlinear system, which cannot be solved analytically in a general case \cite{LL:field}. In such a system various resonant effects take a place \cite{Chirikov:1959, AKN}, and some of them can by described using the adiabatic approach. In particular, a charged particle in the field of an electromagnetic (or electrostatic) wave and a background magnetic field can be trapped in the potential well of the wave and accelerated along the wave front. This phenomenon is called a surfatron acceleration. The mechanism of the surfatron acceleration of charged particles is often considered for description of various plasma-physics phenomena \cite{Sagdeev66,KatsouleasDawson:1983}. Originally this mechanism was suggested for description of charged particles acceleration along the front of a shock wave \cite{Sagdeev66} and this application is still actual \cite{Kichigin:shock}. On the other hand, there are various astrophysical applications of the surfatron acceleration mechanism to problems of generation of high energy particles \cite{Erokhin89, Eliasson, Wang07, Dieckmann05} and consequent radiation \cite{Zaslavskii86, Bulanov86, Bulanov00}. Surfatron acceleration of relativistic particles was considered, for example, in \cite{Chernikov:1992,Itin:2000}.  In all these papers authors consider a  particle interaction with an electrostatic wave.

Surfatron acceleration of a particle by an electromagnetic wave is less studied. The analytical theory was constructed only for nonrelativistic \cite{AVNZ:2009} or ultrarelativistic \cite{Chernikov:1992,Itin:2002} particles. The effect of large particle velocity was estimated numerically in \cite{Takeuchi87,Itin:2002}. Also, numerical calculations were carried out for the case when the wave amplitude is small compared to the background magnetic field \cite{Ginet90, Karimabadi90}; in this case, the particle is accelerated by multiple scatterings on the wave. In addition, several laboratory experiments with relativistic particles and large wave amplitudes were carried out for the investigation of surfatron acceleration of charged particles by electromagnetic waves (see \cite{Yugami96} and references therein). Therefore, a complete analytic theory of relativistic charged particle captures by electromagnetic waves and the resulting acceleration is important.

Particle capture and surfatron acceleration is possible if phase velocity of the wave is smaller than the absolute value of the particle velocity (and hence, smaller than the speed of light). In this case the projection of particle velocity onto the wave vector direction can become equal to value of the phase velocity of the wave and the resonance takes a place. There are several plasma modes that can support a wave with needed properties: the magnetosonic wave with frequency close to lower-hybrid \cite{Neishtadt09}, the plasma wave with frequency close to higher-hybrid \cite{Erokhin89} or various drift modes of plasma instability \cite{Zelenyi09}. In addition, a relatively small population of trapped particles can decrease the phase velocity of a wave \cite{Krasovsky10} and establish the condition of resonance interaction.

A secondary effect of surfatron acceleration is  the radiation of accelerated particles due to the oscillatory component of their motion \cite{Zaslavskii86, Zaslavskii86:jetp}. A captured particle accelerates along the wave front, and at the same time it oscillates near the minimum of the wave potential well. Due to these oscillations the particle can radiate. Estimates of this radiation were carried out in the case of an electrostatic wave. The question of particle radiation in the system with electromagnetic wave is discussed in our paper.

We study the problem of interaction of a charged particle in a uniform magnetic field with an electromagnetic wave using the theory of resonant phenomena.  The study of slow passages of a Hamiltonian system through a nonlinear resonance was started in \cite{Chirikov:1959}. In the present paper we use the theory of resonant processes in Hamiltonian systems with slow and fast motions in the form developed in \cite{Neishtadt:1997,Neishtadt:1999,NV:2006} (see \cite{Neishtadt:1997} for references to preceding works). The description of scattering on resonances and captures into resonances plays a central role in this theory. Resonant phenomena arise in a variety of problems of physics, including hydrodynamics, celestial mechanics, and plasma physics. For several examples of resonant phenomena, see, e.g., \cite{Shklyar:1986,FeingoldandKadanoffandPiro:1988,Gendelman:2001,Mezic:2001,GrosfeldFriedland:2002,VWG:2008,RomKedar:2009}.

\section{ Main equations}\label{main}

We consider motion of a relativistic charged particle of mass $m$ and charge $e$ in a uniform magnetic field ${\bf B}_0 = (0,0,B_0)$  and the field of a plane linearly polarized electromagnetic wave propagating in perpendicular direction to ${\bf B}_0 $. Thus, in the Cartesian coordinates $(\hat q_1,\hat q_2,\hat q_3)$ the resulting magnetic field components are $B_1 = B_2 = 0, \, B_3 = -B_w \sin (\hat k\hat q_1 - \hat \om \hat t) + B_0$, where $B_w$ is the amplitude of the magnetic field of the wave, $\hat k$ is the magnitude of the wave vector directed along the $\hat q_1$-axis, and $\hat \om$ is the wave frequency. The corresponding vector potential can be chosen as
\be
{\bf A} = \left(0, B_0 \hat q_1 + \frac{B_w}{\hat k}\cos(\hat k\hat q_1 - \hat \om \hat t) ,0 \right).
\label{2.1}
\ee
Let $\hat p_i , \, i=1,2,3$ be components of the particle's momentum. Introduce
\be
\calP_2 = \hat p_2 +\frac{e}{c}B_0 \hat q_1 + \frac{eB_w}{c\hat k}\cos(\hat k \hat q_1 - \hat \om \hat t).
\label{2.1a}
\ee
The Hamiltonian of the system is:
\be
\hat H=\sqrt{m^2 c^4 + c^2\hat p_1^2 + \left(c\calP_2 - eB_0\hat q_1 - \frac{eB_w}{\hat k}\cos(\hat k\hat
q_1 - \hat \om \hat t)\right)^2 + c^2\hat p_3^2},
\label{2.2}
\ee
and pairs of canonically conjugate variables are $(\hat p_1,\hat q_1),(\calP_2,\hat q_2),(\hat p_3,\hat q_3)$. The Hamiltonian does not contain variables $\hat q_2$ and $\hat q_3$. Thus canonically conjugate momenta $\calP_2$ and $\hat p_3$ are constants of motion. We can put $\hat p_3 = 0$ (it always can be done by redefining the particle's mass) ; one can also make $\calP_2 = 0$ choosing properly the origin in $\hat q_1$. Introduce Larmor frequency $\om_L = eB_0/(mc)$ and dimensionless parameter $\eps = eB_w/(m\hat kc^2)$. We assume that $\eps$ is small: $0<\eps\ll 1$. Use the following rescaling to make the system dimensionless:
\bea
 p_i = \frac{\hat p_i}{mc}, \;\;  q_i = \frac{\om_L \hat q_i}{\eps c}, \nonumber \\
 t = \frac{\om_L \hat t}{\eps}, \;\;  H = \frac{\hat H}{mc^2}, \nonumber \\
 k = \frac{\hat k c\eps}{\om_L}, \;\;  \om = \frac{\hat \om \eps}{\om_L}. \nonumber
\eea
The Hamiltonian in the new variables is:
\be
H = \sqrt{1+ p_1^2 + (\eps q_1 + \eps \cos(kq_1 - \om t))^2}.
\label{2.3}
\ee
Introduce new variable $U = \om t$. Let $P_U$ be the variable, canonically conjugate to $U$. Thus we obtain a 2 d.o.f. Hamiltonian system. The Hamiltonian takes the form:
\be
\calH = \om P_U + \sqrt{1+ p_1^2 + (\eps q_1 + \eps \cos(kq_1 - U))^2}.
\label{2.4a}
\ee
Now we introduce the wave phase $\ffi = kq_1 - U$ as a new variable. To this end, we make canonical transform with generating function $W = pq_1 + I(kq_1 - U)$, where $I$ is a new variable canonically conjugate to $\ffi$. For the new variables $p,q,I$ we have $q = q_1, \, p = p_1 - Ik, I = -P_U$. Denote $\eps q = \tilde q$. Omitting the tilde, we find for the Hamiltonian in the new variables:
\be
\calH = -\om I + \sqrt{1 + (p+Ik)^2 + (q+\eps \cos \ffi)^2},
\label{2.4}
\ee
where the pairs of canonically conjugate variables are $(I,\ffi)$ and $(p,\eps^{-1} q)$. This Hamiltonian can be represented in the form $\calH = H_0 + \eps H_1$, where
\bea
H_0 = -\om I + \sqrt{1 + (p+Ik)^2 + q^2}, \nonumber \\
\eps H_1 = \eps \frac{q\cos\ffi}{\sqrt{1 + (p+Ik)^2 + q^2}} + O(\eps^2).
\label{2.5}
\eea
In the main approximation, the equations of motion are
\bea
\dot I = -\frac{\pt \calH}{\pt \ffi} = \eps \frac{q\sin\ffi}{\sqrt{1 + (p+Ik)^2 + q^2}}
\label{2.6} \\
\dot \ffi = \frac{\pt \calH}{\pt I} = -\om + \frac{k(p+Ik)}{\sqrt{1 + (p+Ik)^2 + q^2}}
\nonumber \\
\dot p = -\eps \frac{\pt \calH}{\pt q} = -\eps \frac{q}{\sqrt{1 + (p+Ik)^2 + q^2}}
\label{2.7} \\
\dot q = \eps \frac{\pt \calH}{\pt p} = \eps \frac{p+Ik}{\sqrt{1 + (p+Ik)^2 + q^2}}.
\nonumber
\eea
In this system, variable $\ffi$ is fast (its time derivative is a value of order $\eps$), and the other variables are slow (their time derivatives are of order $1$). Thus, one can average over fast phase $\ffi$ and obtain the {\it averaged system}. Motion in this system is just the Larmor rotation in the uniform magnetic field $B_0$. The averaging, however, does not describe the motion adequately near the resonance $\dot\ffi = 0$. At the resonance, projection of particle's velocity onto the $q_1$-axis equals the phase velocity of the wave. Resonance condition $\pt H_0/\pt I = 0$ defines a {\it resonance surface} in the $(p,q,I)$-space:
\be
(p+Ik)^2(k^2 - \om^2) = (1+q^2)\om^2.
\label{2.8}
\ee
We denote the value of $I$ on this surface as $I_r$. This is a function of variables $p,q$:
\be
I_r (p,q) = \frac1{k} \left( \om \sqrt{\frac{1+q^2}{k^2-\om^2}} - p \right).
\label{2.8a}
\ee
Intersection of surface (\ref{2.8}) and isoenergetic surface $H_0 = h$ defines the {\it resonance curve}. Its projection onto the $(p,q)$-plane is a hyperbola given by equation
\be
(h-v_{\phi}p)^2 = (1+q^2)(1-v_{\phi}^2),
\label{2.9}
\ee
where we introduced the dimensionless phase velocity of the wave $v_{\phi} = \om/k$. Note that $v_{\phi}$ is always smaller than 1.

Variable $I$ is an integral of motion of the averaged system (see (\ref{2.6})) and thus an adiabatic invariant of the exact system. Far from the resonance the value of $I$ is preserved with the accuracy of order $\eps$ on time intervals of order $1/\eps$ (see, e.g., \cite{AKN}). The adiabatic invariance of $I$ breaks down near the resonance, where the averaging does not work properly. In a neighborhood of the resonance phenomena of {\it capture} and  {\it scattering} can take place. We study dynamics of the system near the resonance in the following sections.

\section{ Motion near the resonance}\label{resonance}

To study the system near the resonance, we apply the approach formulated in \cite{Neishtadt:1999} (see also \cite{NV:2006}). Close to the surface (\ref{2.8}) the Hamiltonian can be expanded into series in $(I-I_r)$:
\be
\calH =\La(p,q) +\frac12 g(p,q)(I-I_r(p,q))^2 + \eps H_1|_{I=I_r} + O(|I-I_r|^3) + O(\eps(I-I_r(p,q))^2).
\label{3.1}
\ee
Here $\La(p,q) = H_0|_{I=I_r}$ is the unperturbed Hamiltonian $H_0$ restricted onto the resonant surface (\ref{2.8}). Function $g(p,q)$ in (\ref{3.1}) is $\pt^2 H_0/\pt I^2$ restricted onto the resonant surface.  It is straightforward to find
\be
\La(p,q) = pv_{\phi} + \sqrt{(1+q^2)(1-v_{\phi}^2)}, \;\; g(p,q) = k^2
\frac{(1-v_{\phi}^2)^{3/2}}{\sqrt{1+q^2}}.
\label{3.2}
\ee
Introduce notation $d = q\sqrt{1-v_{\phi}^2}/\sqrt{1+q^2}$. Then
\be
\eps H_1|_{I=I_r} = \eps d \cos\ffi.
\label{3.3}
\ee
Now we introduce new momentum $K=I - I_r(p,q) + O(\eps)$. To this end, we make a canonical transformation of variables $(I,\ffi,p,q)\mapsto (K,\bar\ffi,\bar p,\bar q)$ with generating function $W_1 = \bar p \eps^{-1} q +(K + I_r(\bar p,q))\ffi$. Omitting bars, we find for the Hamiltonian in the new variables (we neglect terms of higher orders):
\be
\calH = \La(p,q) + \frac12 g(p,q)K^2 +\eps d \cos\ffi +\eps b(p,q)\ffi,
\label{3.4}
\ee
where
\be
b(p,q) = \{I_r,\La\} = \frac{\pt I_r}{\pt q}\frac{\pt \La}{\pt p}-\frac{\pt I_r}{\pt p}\frac{\pt \La}{\pt
q}= \frac{q}{k\sqrt{1+q^2}}\frac1{\sqrt{1-v_{\phi}^2}}.
\label{3.5}
\ee
Introduce $P = K/\sqrt\eps$, $\theta = t\sqrt\eps$, and rescaled Hamiltonian $F = \calH/\eps$. The rescaled system is Hamiltonian, and pairs of canonically conjugate variables are $(P,\ffi)$ and $(p,\eps^{-3/2}q)$. In the main approximation, the Hamiltonian is
\be
F=\eps^{-1}\La(p,q) + F_0(P,\ffi,p,q), \;\; F_0 = \frac12 gP^2 + d\cos\ffi +b\ffi,
\label{3.6}
\ee
and the equations of motion are
\bea
p^\prime = -\sqrt\eps \frac{\pt \La}{\pt q}, \;\; q^\prime = \sqrt\eps \frac{\pt \La}{\pt p},
\nonumber\\
P^\prime = - \frac{\pt F_0}{\pt \ffi}, \;\; \ffi^\prime = \frac{\pt F_0}{\pt P},
\label{3.7}
\eea
where prime denotes derivative over $\theta$. One can see that variables $(P,\ffi)$ are fast, and variables $(p,q)$ are slow. Thus, as the first step to study this system, one can consider variation of the fast variables at fixed values of $p$ and $q$. Dynamics of the fast variables is defined by Hamiltonian $F_0$, which contains $q$ as a parameter. This is a Hamiltonian of a pendulum under the action of external torque. Consider phase portrait of this system (fast subsystem).  If $d/b > 1$, there is a separatrix on the phase portrait, and, correspondingly, domain of oscillatory motion (see Fig.1). In the opposite case $d/b < 1$, there is no separatrix. Note that ratio $d/b = k(1-v_\phi^2)$ is independent of $q$. In original dimensional units $d/b = (B_w/B_0)(1-\hat v_\phi^2/c^2)$, where $\hat v_{\phi}=\hat \om/\hat k$.  We consider first the case $d/b>1$.

\begin{figure}[t]
\center\epsfig{file=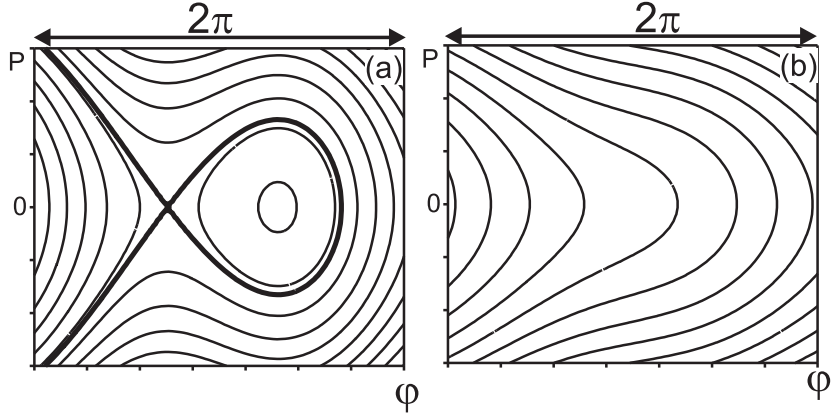, width=0.8\textwidth}
\caption{\label{schem} Schematic view of the phase portrait of the system (\ref{3.6}) for $d/b>1$ (a) and
for $d/b<1$ (b)}
\end{figure}

It is straightforward to obtain for the area $S$ inside the separatrix on the plane $(\ffi,P)$
\be
S = 2\int_{\ffi_s}^{\ffi_m}\left(2\frac{d}{g}\left[\cos\ffi_s + \frac{b}{d}\ffi_s - \cos \ffi -
\frac{b}{d}\ffi\right]\right)^{1/2} \dd \ffi,
\label{3.8}
\ee
where $\ffi_s = \arcsin(b/d)$ and $\ffi_m$ is the root of equation $\cos\ffi +(b/d)\ffi = \cos\ffi_s +(b/d)\ffi_s$ different from $\ffi_s$. One can see that both integration limits and the expression in square brackets in (\ref{3.8}) do not depend on $q$. Thus,
\be
S=\sqrt{\frac{d}{g}}A = \sqrt{\frac{q}{k^2 - \om^2}}A,
\label{3.9}
\ee
where $A$ is a constant independent of $q$ (and $p$).

Now take into account slow variation of $p$ and $q$ according to the first two equations in (\ref{3.7}). It follows from the expression (\ref{3.2}) for $\La(p,q)$ that if $v_\phi>0$, variable $q$ grows with time. This means that area $S$ also grows in the process of motion. Therefore, phase points on the phase portrait of the fast subsystem can cross the separatrix and enter the domain of oscillations. This corresponds to a capture into resonance. The area $J$ encircled by a trajectory in the domain of oscillations on this portrait is an adiabatic invariant (it is called the {\it inner adiabatic invariant}). Thus, as $S$ monotonously grows with time, a captured particle goes deeper and deeper inside the oscillation domain and cannot leave it. This means that a particle captured into the resonance is captured forever.

\begin{figure}[t]
\center\begin{tabular}{cc} \hspace*{-10mm}\epsfig{file=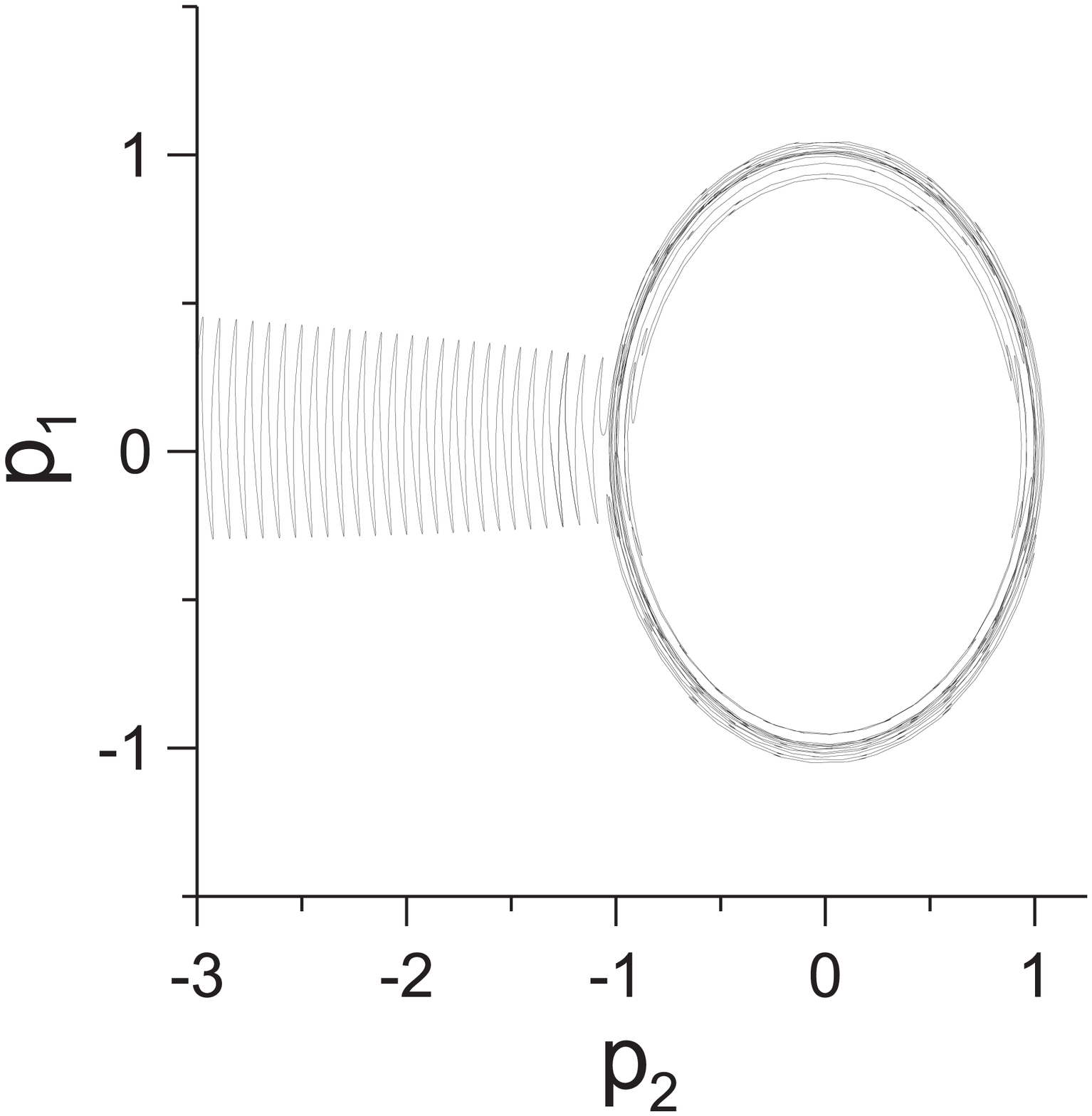,width=2.8in}&
\epsfig{file=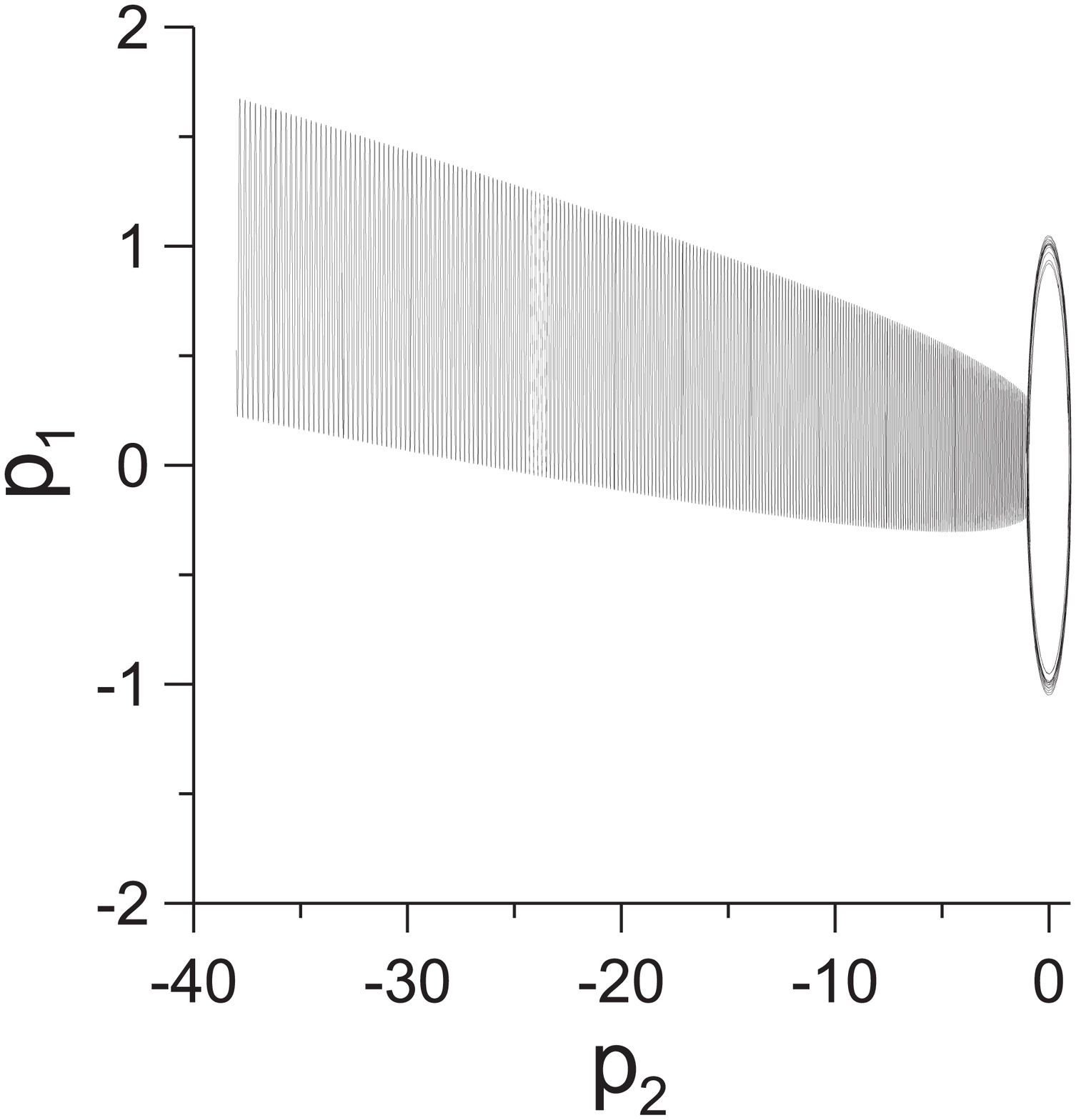,width=2.8in}
\end{tabular}
  \caption{\label{Fig1}Left panel: trajectory of the particle in momentum space $(p_1,p_2)$, Right panel:
the same situation as in left panel with a longer time range. }
\end{figure}

Motion of a captured particle can be described as follows. In the main approximation, it moves with the {\it resonant flow} defined by Hamiltonian $\La(p,q)$. The corresponding  equations of motion are:
\bea
\dot p = -\eps \frac{\pt \La}{\pt q}= \eps q \,\sqrt{\frac{1-v_{\phi}^2}{1+q^2}},
\nonumber\\
\dot q = \eps \frac{\pt \La}{\pt p}= \eps v_{\phi}.
\label{3.10}
\eea
It means that a captured particle moves in $\hat q_1$-direction with the wave at a speed of the wave's phase velocity. It follows from (\ref{2.1a}) and the fact that $\calP_2 = 0$ that  $p_2 \approx -q$, and, hence $p_2 \sim -\eps v_{\phi} t$. Therefore  $\hat p_2$-component of the particle's momentum varies (on average) linearly in time:
$$
\hat p_2 \sim -\frac{e}{c} B_0 \hat v_{\phi}\hat t \sim mc(\omega_L \hat v_{\phi} \hat t/c)
$$

Thus, the particle is accelerated along the wave front. This acceleration is called surfatron one. To find variation of $p_1$ in this motion, we use that $p_1 = p +Ik$ and expression (\ref{2.8a}) for $I$ on the resonant surface. Thus we obtain $p_1 = v_{\phi}\sqrt{(1+q^2)/(1-v_{\phi}^2)}$ and
\be
\dot p_1 = \frac{\eps v_{\phi}^2}{\sqrt{1-v_{\phi}^2}} \frac{q}{\sqrt{1+ q^2}},
\label{3.11}
\ee
where we used the second equation in (\ref{3.10}). At $q \gg 1$ we find that $p_1$ grows with time as $\eps v_{\phi}^2 t/\sqrt{1-v_{\phi}^2}$. In dimensional variables, we find that
$$
 \hat p_1 \sim \frac{eB_0 \hat v_{\phi}^2}{c^2 \sqrt{1 - \hat v_{\phi}^2/c^2}} \hat t \sim mc \frac{ \hat
v_{\phi}/c}{\sqrt{1 - \hat v_{\phi}^2/c^2}} (\omega_L \hat v_{\phi} \hat t/c).
$$
For the energy of a captured particle $E= \sqrt{1 + p_1^2 +p_2^2}$ we find that it also grows linearly with time at large enough values of $q$. Namely, we have $E \sim \eps v_{\phi} t/ \sqrt{1 - v_{\phi}^2}$ and, in dimensional variables, $\hat E = mc^2 E \sim mc \hat v_{\phi} \om_L \hat t/  \sqrt{1 - v_{\phi}^2/c^2}$.

\begin{figure}[t]
\center\epsfig{file=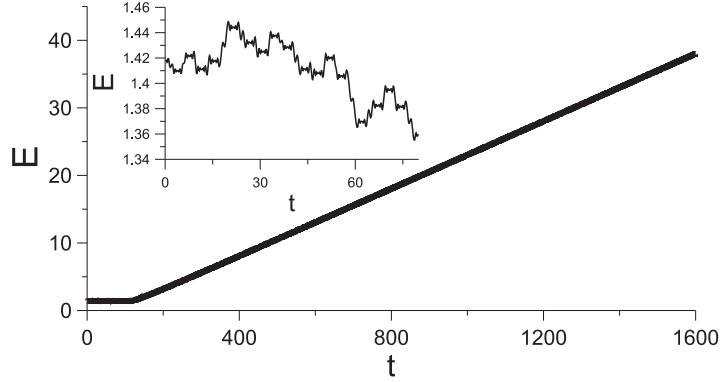, width=0.7\textwidth}
\caption{\label{Fig3} Particle's energy  as a function of time. The inner panel shows the short time interval before
capture.}
\end{figure}

The captured particle also oscillates in the potential well of the wave. These oscillations correspond to motion in the oscillatory domain in Fig. 1. One can evaluate the amplitude and the frequency of the oscillations using conservation of the inner adiabatic invariant $J$. For a captured particle $J$ equals the area inside the separatrix {\it at the time when the particle crossed the separatrix}. It follows from the expression for $F_0$ in (\ref{3.6}) that for a captured particle
\be
J = \frac{2^{3/2}}{k} \,\sqrt{\frac{q}{1-v_{\phi}^2}} \int_{\ffi_1}^{\ffi_2}\sqrt{\tilde f_0 - \cos\ffi -
\frac{\ffi}{k(1-v_{\phi}^2)}}\, \dd\ffi,
\label{3.12}
\ee
where $\tilde f_0 = F_0/d$ does not depend on $\ffi$; $\ffi_1$ and $\ffi_2$ are the minimal and the maximal values of $\ffi$ on a trajectory with fixed values of $F_0$ and $q$. Thus, growth of $q$ results in decreasing of the amplitude of the $\ffi$-oscillations. When the amplitude of these oscillations is sufficiently small, one can expand the Hamiltonian $F_0$ to obtain $2(F_0 - \bar F_0) \approx gP^2 + d |\cos \ffi_0| \cdot (\ffi -\ffi_0)^2$, where $\bar F_0$ and $\ffi_0$ are values of $F_0$ and $\ffi$ at the bottom of the potential well inside the separatrix (see Fig. 1a). The frequency of oscillations (in terms of rescaled time $\theta$) is approximately $\sqrt{gd|\cos \ffi_0| }$, and in this approximation $2\pi J \sim (F_0 - \bar F_0)/\sqrt{gd}$. When the particle is captured, $J = J_0$. Using conservation of $J$ along the trajectory of the captured particle, one obtains the scalings $\Dt\ffi \sim q^{-1/4} (k^2-\omega^2)^{1/4} $ and $\Dt P \sim q^{1/4}(k^2-\omega^2)^{-1/4}$, where $\Dt\ffi$ and $\Dt P$ are amplitudes of $\ffi$-oscillations and $P$-oscillations accordingly. Thus, the amplitude of oscillations in $q$ decreases with time proportionally to $t^{-1/4}$, while amplitude of oscillations in $p_1$ grows with time proportionally to $t^{1/4}$. Accordingly, amplitude of oscillations in $p_2$ decreases with time as $t^{-1/4}$.  In dimensional variables we have $\Dt\hat q_1 \sim \eps ( \hat v_{\phi} \hat t)^{-1/4}$ and $\Dt \hat p_1 \sim \sqrt{\eps} ( \hat v_{\phi} \hat t)^{-1/4}$.  The frequency of these oscillations is $\om_0 \approx \sqrt{\eps gd}$ (we recall that we made time rescaling to obtain (\ref{3.6})). Thus,
\be
\om_0 \approx k(1-v_{\phi}^2) \sqrt{\frac{\eps q}{1+q^2}} .
\label{3.13}
\ee
At $q \gg 1$ we find that this frequency decreases as $\sqrt{\eps/q}$. In dimensional variables we find that $\hat \om_0 \sim (\eps \hat t)^{-1/2}$.

\begin{figure}[t]
\epsfig{file=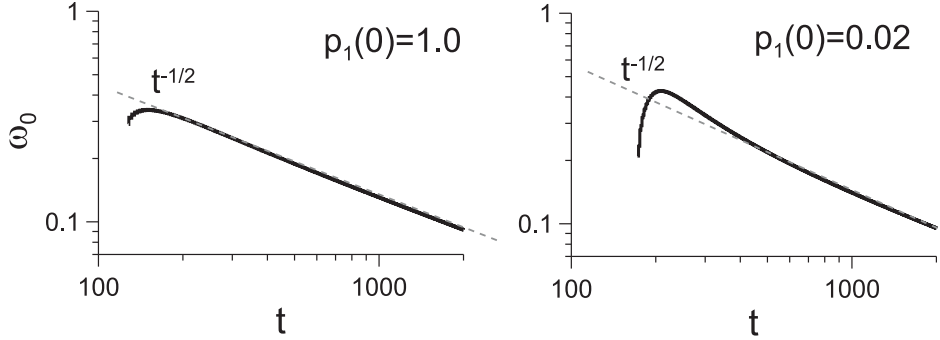, width=5.0in}
\caption{\label{Fig4} Oscillation frequency of $\dot \varphi$ as a function of time for two particles with different
values of initial momentum. Grey dashed lines show the theoretical dependence $\omega_0 \sim t^{-1/2}$}
\end{figure}

Here are more complete formulas in dimensional variables for the amplitudes of the oscillations at large enough values of $\hat q_1$, such that we can put $\hat q_1 = \hat v_{\phi} \hat t$. One obtains:
$$
\Dt \hat q_1 \sim \varepsilon c/\om_L \sqrt{B_0/B_w} \left(1 - \frac{ \hat v_{\phi}^2}{c^2}\right)^{1/4}(
\om_L \hat v_{\phi} \hat t/c)^{-1/4},
$$
$$
\Dt \hat p_1 \sim \varepsilon^{1/2} \sqrt{B_w/B_0} \left(1 - \frac{ \hat v_{\phi}^2}{c^2}\right)^{1/4} (
\om_L \hat v_{\phi} \hat t/c)^{1/4} ,
$$
$$
\hat\om_0 \sim \omega_L \varepsilon^{-1/2} \sqrt{B_w/B_0} \left(1 - \frac{ \hat v_{\phi}^2}{c^2}\right) (
\om_L \hat v_{\phi} \hat t/c)^{-1/2}.
$$

Capture into the resonance is a probabilistic phenomenon (see, e.g., \cite{Neishtadt:1999,NV:2006}). Its probability is a small value of order $\sqrt\eps$. However, the geometry of the system makes the particle pass through the resonance repeatedly, at each Larmor turn. The probability of capture after $\sim\eps^{-1/2}$ Larmor turns is a value of order one (provided that $d/b > 1$).

For comparison with theoretical results we present the numerical solution of the system with Hamiltonian (\ref{2.3}), parameters $\varepsilon=0.1$, $\hat \omega/(\hat kc)=0.25$ and initial value of the $p_1 = 0.1$. The particle trajectory in momentum space $(p_1, p_2)$ is shown in Fig. \ref{Fig1}. At the first stage of modelling the particle rotates in the constant background magnetic field (Larmor rotation). This motion is slightly perturbed by influence of the wave: the Larmor circles in $(p_1, p_2)$ plane are ``scattered''.  Then after certain time interval the particle is captured by the wave and the magnitude of momentum $p_2$ grows with time while momentum $p_1$ oscillates around the resonant value (it is also increasing, yet much slower). To compare the scale of growth of momenta $p_1$ and $p_2$ we plot the same picture on a longer time range  (see Fig. \ref{Fig1}). The relation between $p_1$ and $p_2$ after initial time interval is $1/40 \sim \varepsilon\hat \omega/(\hat kc)$, in agreement with the  theory (see equation (\ref{3.11})). The particle energy $E = \sqrt{1+p_1^2+p_2^2}$ is shown in Fig. \ref{Fig3}. The energy is almost constant before capture (if we neglect small scatterings due to wave impacts) and after the capture the energy grows linearly with time. In addition we examine the theoretical equation for frequency of oscillation of the captured particle - equation (\ref{3.13}). For this purpose we plot the oscillation frequency of $\dot \varphi$ around the null value - Fig.~\ref{Fig4}.


\section{Scatterings on the resonance}

Capture into the resonance is impossible in the case $d/b \le 1$, when there is no oscillatory domain on the phase portrait of the pendulum-like system (see Fig. 1b).  However, in this case the particle energy also changes at the resonance crossing. This happens due to scatterings on the resonance. If $d/b \le 1$, the average value of the jump in the energy is zero (see \cite{Neishtadt:1999}), but the dispersion is non-zero, and thus diffusive variation of the particle energy may be possible. Here we study this topic more attentively.

When the particle is far from the resonance, its energy  is approximately constant, because the impact of the wave can be averaged. Thus, to study variation of the particle energy we find its time derivative according to equations of motion (\ref{3.7}) and integrate it near the resonance. We have
\be
E=\sqrt{1+p_1^2 + q^2} = \sqrt{1+(p+kI)^2 + q^2}.
\label{4.1}
\ee
Using (\ref{2.4}) we can write
\be
\dot E = \frac{\dd }{\dd t} (\calH + \om I) = \om \dot I = -\om \frac{\pt \calH}{\pt \ffi}.
\label{4.2}
\ee
From (\ref{3.1}) and (\ref{3.3}) we find that near the resonance
\be
\dot E = \eps \om \, d \sin \ffi ,
\label{4.3}
\ee
To integrate (\ref{4.3}) we change the integration variable from time $t$ to phase $\ffi$ according to $\dot \ffi = \sqrt{\eps} \pt F_0/\pt P$. Thus we find for variation (jump) of the particle energy on one resonance crossing
\be
j_E = 2\sqrt{\eps} \frac{v_\phi}{\sqrt{1-v_\phi^2}} \sqrt{q} \int_{-\infty}^{\ffi_*} \frac{\sin \ffi \,
\dd\ffi}{\sqrt{2\left[\cos\ffi_* + \frac{b}{d}\ffi_* - \cos \ffi - \frac{b}{d}\ffi\right]}},
\label{4.4}
\ee
where $\ffi_*$ is the wave's phase at the resonance crossing, and $q$ is taken also at the crossing of the unperturbed trajectory with energy $E$ and the resonant surface. The value of $\ffi_*$ strongly depends on initial conditions and should be treated as random. Therefore, change in the particle's energy on the resonance is also a random variable. If there is no separatrix on the phase portrait in Fig. 1 (i.e., if $d/b \le 1$), the average value of this latter random value is zero. An important question is whether these values at successive crossings are statistically independent. Expressing $q$ in (\ref{4.4}) on the resonance via $E$ (we use that at the resonance $E = \sqrt{(1+q^2)/(1-v_\phi^2)}$) we find that at $E \gg 1$ the variation of energy at the resonance scattering is a value of order
\be
j_E \sim \sqrt{\eps E}.
\label{4.5}
\ee

On the plane $(q,p_1)$ unperturbed motion of the particle is rotation along the circle $p_1^2 + q^2 = E^2 - 1$. Using Hamiltonian equations of the unperturbed motion, one immediately finds that the frequency of this rotation is $\om_0 = \eps/E $. The resonant curve on the plane $(q,p_1)$ is a branch of hyperbola $p_1^2 (1-v_\phi^2) - q^2 v_\phi^2= v_\phi^2$ with $p_1>0$. At large enough values of $E$ the trajectory of the unperturbed motion crosses the resonant curve at two points. It is straightforward to find that the time of motion between these two points is a value of order $E/\eps$. Consider two successive resonance crossings. Let the values of $\ffi$ at the first and the second crossings be $\ffi_*^{(1)}$ and $\ffi_*^{(2)}$ accordingly. To find $\ffi_*^{(2)} - \ffi_*^{(1)}$ one can integrate equation of motion for $\ffi$ in (\ref{2.6}). Thus, one obtains $\ffi_*^{(2)} - \ffi_*^{(1)} \sim E/\eps$.  A small variation $\dt\ffi_*^{(1)}$ of the phase $\ffi_*^{(1)}$ results in variation of the energy jump $\dt j_E \sim \sqrt{\eps E}\dt\ffi_*^{(1)}$ at  the first resonance crossing. The resulting variation $\dt \ffi_*^{(2)}$ of the phase at the second crossing can be found as $\dt j_E\, \pt(\ffi_*^{(2)} - \ffi_*^{(1)})/\pt E \sim \dt j_E/\eps \sim \dt\ffi_*^{(1)}\sqrt{E}/\sqrt{\eps}$. Thus, the resulting variation in the phase at the second resonance crossing is much larger than $\dt\ffi_*^{(1)}$. Therefore, the values of phase at successive resonance crossings are statistically independent. Hence, the jumps in the particle's energy at the resonance produce diffusive variation of the energy and its unlimited stochastic growth. Note, that in \cite{AVNZ:2009} the diffusive growth of energy was studied in nonrelativistic case. It was found that, unlike in the relativistic case,  for a nonrelativistic particle the energy diffusion slows down and finally stops at large enough energies.

\begin{figure}[t]
\center\epsfig{file=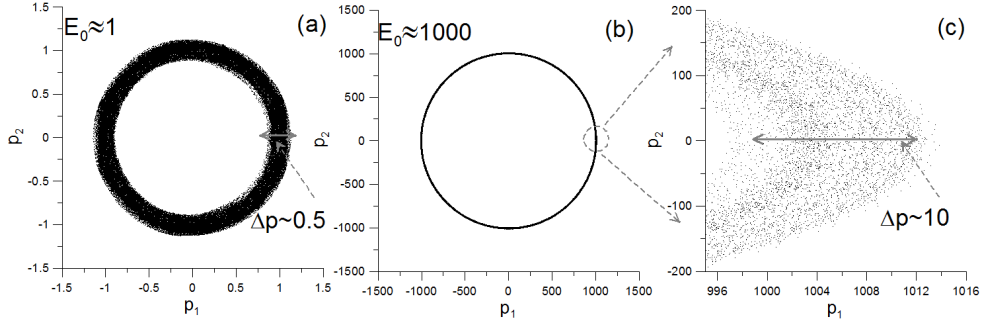, width=0.95\textwidth}
\caption{\label{dif} The Poincar\'e sections for several cases with different initial energy ($\varepsilon
= 0.1$). Number of resonance crossings in both cases is $10^6$. The energy diffusion is clearly seen.}
\end{figure}

One can estimate the rate of the energy diffusion as follows. Consider a long trajectory that crosses the resonance $N \gg 1$ times. Introduce new variable $\kappa = \sqrt E$. It follows from (\ref{4.5}) that at every resonant crossing $\kappa$ changes by a value of order $\sqrt\eps$. If successive jumps in energy are not correlated, it follows from (\ref{4.5}) that typical displacement of $\kappa$ after $N$ resonance crossings is $\Dt\kappa \sim \sqrt{\eps N}$. Hence, typical value of energy after $N$ jumps is proportional to $N$:
\be
E \sim \eps N.
\label{4.6}
\ee
Time interval between successive jumps is a value of order of the Larmor period. Hence, the time interval corresponding to $N$ resonance crossings is $t \sim N E/\eps$. Combining this with (\ref{4.6}), we find that the energy typically grows with time as
\be
E  \sim \eps \sqrt{t}.
\label{4.7}
\ee
We also obtain that the number of jumps (resonance crossings) grows with time as $N \sim \sqrt{t}$.

\begin{figure}[t]
\center\epsfig{file=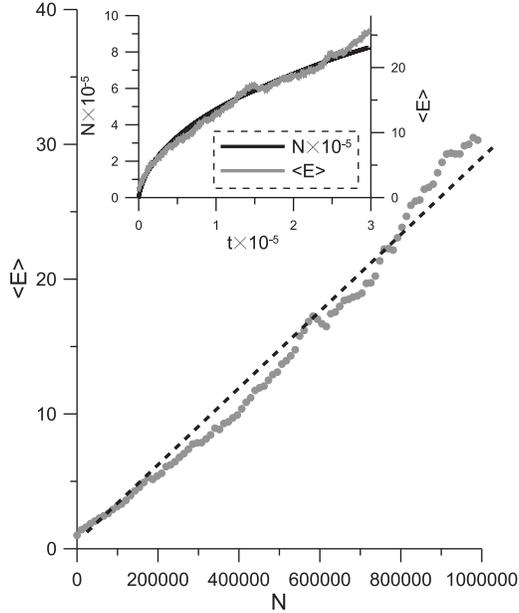, width=0.5\textwidth}
\caption{\label{dif2} The energy of particle ensemble (number of particles is $10^3$) as a function of
number of resonance crossings in the system with $\varepsilon = 0.1$. In subpanel: the number of resonance
crossings and averaged energy as a function of time.}
\end{figure}

These results on the energy diffusion of a relativistic particle can be examined numerically. For this purpose we construct the Poincar\'e section of a particle trajectory in $(p_1, p_2)$ plane. Points on this plane are plotted with time period $2\pi/\omega$. One can see that the diffusion in $(p_1, p_2)$ space become stronger as the initial energy of particles grows (Fig. \ref{dif}). In Fig. \ref{dif2}, we present results of numerics illustrating estimates (\ref{4.6}) and (\ref{4.7}).

\section{Energy loss due to radiation}

The oscillations of the captured particle across the wave front result in energy loss due to radiation. The energy loss has the following rate (see \cite{LL:field}):
\be
\frac{\dd \hat E_-}{\dd \hat t} = \frac{2e^2 }{3c^3 }\left(\frac{\dd^2\hat q_1}{\dd \hat t^2}\right)^2.
\label{5.1}
\ee
In the dimensionless variables this expression takes the form:
\be
\frac{\dd E_-}{\dd t} = \frac{2r_0 \om_L}{3 \eps^3 c}\left(\frac{\dd^2 q}{\dd t^2}\right)^2,
\label{5.2}
\ee
where we introduced notation $r_0 =e^2/mc^2$. For a captured particle moving deep inside the domain of oscillations on phase portrait in Fig.~1b, we find
\be
\frac{\dd^2 q}{\dd t^2} = -\eps k^{-1} \Dt\ffi \om_0^2 \sin\om_0 t = K_1 \frac{\eps^2 q^{3/4}}{1+q^2} k^{3/2}(1-v_{\phi}^2)^{9/4} \sin\om_0 t.
\label{5.3}
\ee
Here $\Dt\ffi$ and $\om_0$ are amplitude and frequency of small oscillations of the captured particle, and $K_1$ is a value of order one; $K_1$ depends on the value of $q$ at the instance of capture into the resonance (see Section 3). Thus, we find
\be
\frac{\dd E_-}{\dd t} = \eps K_2 \frac{r_0 \om_L}{  c} k^3  \frac{q^{3/2}}{(1+q^2)^2} (1-v_{\phi}^2)^{9/2}.
\label{5.4}
\ee
Here $K_2$ is also a value of order one. In the latter expression, the fraction containing $q$ reaches its maximum value $f_{max} \approx 0.266$ at $q = \sqrt{3/5}$. Hence, loss of energy due to radiation is maximal at this value of $q$.

On the other hand, the captured particle is accelerated along the wave front, and thus it gains energy at the rate (see Section 3)
\be
\frac{\dd E_+}{\dd t} = \eps v_{\phi}/\sqrt{1-v_{\phi}^2}.
\label{5.5}
\ee
Comparing expressions (\ref{5.4}) and (\ref{5.5}) we find that if
\be
K_2 f_{max} \frac{r_0}{cv_{\phi}} \om_L k^3(1-v_{\phi}^2)^{5} < 1,
\label{5.6}
\ee
the radiation cannot stop the particle acceleration. In the opposite case,
\be
K_2 f_{max} \frac{r_0}{cv_{\phi}} \om_L k^3(1-v_{\phi}^2)^{5} > 1,
\label{5.7}
\ee
the acceleration can be stopped under the additional condition that the capture took place at a value of $q$ smaller than the largest of the two roots of equation $\dd E_+/\dd t =\dd E_-/\dd t$.

Note that inequality (\ref{5.7}) can be written in dimensional form as $ r_0 \omega _L \left( B_w /B_0  \right)^3 \left( {1 - \hat v_\phi ^2 /c^2 } \right)^5  > \hat v_\phi $. To be valid, this inequality needs either $B_w$ very large or $v_{\phi}$ very small. Neither so large value of magnetic field, nor so small value of phase velocity can be found in physically realistic situations, and hence (\ref{5.7}) cannot be valid in such situations.

\section{Conclusions}

In this paper we considered dynamics of a relativistic charged particle in the field of an electromagnetic wave in the presence of a background magnetic field. We have described the particle capture into resonance with the wave and consequent acceleration using approach of the adiabatic theory of motion. During the acceleration the particle's momentum in the direction of the wave vector $\hat p_1$ and along the wave front $\hat p_2$ change with time linearly ($\hat p_1 \sim mc \frac{ \hat v_{\phi}/c}{\sqrt{1 - \hat v_{\phi}^2/c^2}} (\omega_L \hat v_{\phi} \hat t/c) $, $\hat p_2 \sim -mc(\omega_L \hat v_{\phi} \hat t/c)$ ). As a result the particle energy $\hat E \sim  \sqrt{m^2c^4+\hat p_1^2+\hat p_2^2}$ grows with time as $\hat E \sim mc^2(\omega_L \hat v_{\phi} \hat t/c)$. The estimates of energy loss due to radiation of the accelerated particle show that it is not sufficient to stop the acceleration and the particle energy grows infinitely. If the condition of capture into the resonance is not satisfied (the magnitude of the wave is less than a certain value), particle can nevertheless gain energy by scatterings on the resonance. In this case $\hat E \sim N$, where $N$ is a number of resonance crossings.

\section*{Acknowledgements}

The work was supported in part by the Russian Foundation for Basic Research (project nos. 09-01-00333, 08-02-00201), and the Council of the Russian Federation Presidential Grants for State Support of Leading Scientific Schools (project no. NSh-8784.2010.1).


\end{document}